\documentclass[aps,prl,twocolumn,superscriptaddress]{revtex4-2}

\usepackage{graphicx}
\usepackage{dcolumn}
\usepackage{bm}
\usepackage{rotating}

\usepackage{bm,color}

\usepackage{tabularx}
\usepackage{epstopdf}

\usepackage{ulem}
\usepackage{hyperref}
\usepackage{tikz-feynman}

\usepackage{multirow}

\usepackage{overpic}
\usepackage{verbatim}

\usepackage{lineno}

\newcommand{\jpc}{J^{PC}}
\newcommand{\pip}{\pi^+}
\newcommand{\pim}{\pi^-}
\newcommand{\piz}{\pi^0}

\newcommand{\etac}{\eta_c}

\newcommand{\hc}{h_c}
\newcommand{\jpsi}{J/\psi}
\newcommand{\ks}{K_{S}^{0}}
\newcommand{\EE}{e^+e^-}
\newcommand{\MM}{\mu^+\mu^-}
\newcommand{\pp}{\pi^+\pi^-}
\newcommand{\kk}{K^+K^-}

\newcommand{\ppbar}{p\bar{p}}
\newcommand{\pipikk}{\pi^{+}\pi^{-}K^{+}K^{-}}
\newcommand{\pipippbar}{\pi^{+}\pi^{-}p\bar{p}}
\newcommand{\kkkk}{2(K^{+}K^{-})}
\newcommand{\pipipipi}{2(\pi^+\pi^-)}
\newcommand{\pipipipipipi}{3(\pi^+\pi^-)}
\newcommand{\pipipipikk}{2(\pi^+\pi^-)K^{+}K^{-}}
\newcommand{\kskpi}{K^{0}_S K^{\pm}\pi^{\mp}}
\newcommand{\kskpipipi}{K^{0}_{S} K^{\pm}\pi^{\mp}\pi^{+}\pi^{-}}
\newcommand{\kkpiz}{\kk\piz}
\newcommand{\ppbarpiz}{\ppbar\piz}
\newcommand{\kketa}{\kk\eta}
\newcommand{\pipieta}{\pp\eta}
\newcommand{\pipipizpiz}{\pp\piz\piz}
\newcommand{\pipipipieta}{2(\pp)\eta}
\newcommand{\pipipipipizpiz}{2(\pp\piz)}

\newcommand{\mev}{\rm{MeV}}

\newcommand{\mevcc}{\rm{MeV}/c^2}
\newcommand{\gev}{\rm{GeV}}

\begin{document}

\title{\boldmath
Observation of Three Resonant Structures
in the Cross Section of $e^+e^-\to\pi^+\pi^- h_c$
}

\author{
M.~Ablikim$^{1}$, M.~N.~Achasov$^{4,c}$, P.~Adlarson$^{77}$, X.~C.~Ai$^{82}$, R.~Aliberti$^{36}$, A.~Amoroso$^{76A,76C}$, Q.~An$^{73,59,a}$, Y.~Bai$^{58}$, O.~Bakina$^{37}$, Y.~Ban$^{47,h}$
% ...
\\
\vspace{0.2cm}
(BESIII Collaboration)\\
\vspace{0.2cm} {\it
$^{1}$ Institute of High Energy Physics, Beijing 100049, People's Republic of China\\
$^{2}$ Beihang University, Beijing 100191, People's Republic of China\\
$^{3}$ Bochum  Ruhr-University, D-44780 Bochum, Germany\\
$^{4}$ Budker Institute of Nuclear Physics SB RAS (BINP), Novosibirsk 630090, Russia\\
$^{5}$ Carnegie Mellon University, Pittsburgh, Pennsylvania 15213, USA\\
$^{6}$ Central China Normal University, Wuhan 430079, People's Republic of China\\
$^{7}$ Central South University, Changsha 410083, People's Republic of China\\
$^{8}$ China Center of Advanced Science and Technology, Beijing 100190, People's Republic of China\\
$^{9}$ China University of Geosciences, Wuhan 430074, People's Republic of China\\
$^{10}$ Chung-Ang University, Seoul, 06974, Republic of Korea\\
$^{11}$ COMSATS University Islamabad, Lahore Campus, Defence Road, Off Raiwind Road, 54000 Lahore, Pakistan\\
$^{12}$ Fudan University, Shanghai 200433, People's Republic of China\\
$^{13}$ GSI Helmholtzcentre for Heavy Ion Research GmbH, D-64291 Darmstadt, Germany\\
$^{14}$ Guangxi Normal University, Guilin 541004, People's Republic of China\\
$^{15}$ Guangxi University, Nanning 530004, People's Republic of China\\
$^{16}$ Guangxi University of Science and Technology, Liuzhou 545006, People's Republic of China\\
$^{17}$ Hangzhou Normal University, Hangzhou 310036, People's Republic of China\\
$^{18}$ Hebei University, Baoding 071002, People's Republic of China\\
$^{19}$ Helmholtz Institute Mainz, Staudinger Weg 18, D-55099 Mainz, Germany\\
$^{20}$ Henan Normal University, Xinxiang 453007, People's Republic of China\\
$^{21}$ Henan University, Kaifeng 475004, People's Republic of China\\
$^{22}$ Henan University of Science and Technology, Luoyang 471003, People's Republic of China\\
$^{23}$ Henan University of Technology, Zhengzhou 450001, People's Republic of China\\
$^{24}$ Huangshan College, Huangshan  245000, People's Republic of China\\
$^{25}$ Hunan Normal University, Changsha 410081, People's Republic of China\\
$^{26}$ Hunan University, Changsha 410082, People's Republic of China\\
$^{27}$ Indian Institute of Technology Madras, Chennai 600036, India\\
$^{28}$ Indiana University, Bloomington, Indiana 47405, USA\\
$^{29}$ INFN Laboratori Nazionali di Frascati , (A)INFN Laboratori Nazionali di Frascati, I-00044, Frascati, Italy; (B)INFN Sezione di  Perugia, I-06100, Perugia, Italy; (C)University of Perugia, I-06100, Perugia, Italy\\
$^{30}$ INFN Sezione di Ferrara, (A)INFN Sezione di Ferrara, I-44122, Ferrara, Italy; (B)University of Ferrara,  I-44122, Ferrara, Italy\\
$^{31}$ Inner Mongolia University, Hohhot 010021, People's Republic of China\\
$^{32}$ Institute of Modern Physics, Lanzhou 730000, People's Republic of China\\
$^{33}$ Institute of Physics and Technology, Mongolian Academy of Sciences, Peace Avenue 54B, Ulaanbaatar 13330, Mongolia\\
$^{34}$ Instituto de Alta Investigaci\'on, Universidad de Tarapac\'a, Casilla 7D, Arica 1000000, Chile\\
$^{35}$ Jilin University, Changchun 130012, People's Republic of China\\
$^{36}$ Johannes Gutenberg University of Mainz, Johann-Joachim-Becher-Weg 45, D-55099 Mainz, Germany\\
$^{37}$ Joint Institute for Nuclear Research, 141980 Dubna, Moscow region, Russia\\
$^{38}$ Justus-Liebig-Universitaet Giessen, II. Physikalisches Institut, Heinrich-Buff-Ring 16, D-35392 Giessen, Germany\\
$^{39}$ Lanzhou University, Lanzhou 730000, People's Republic of China\\
$^{40}$ Liaoning Normal University, Dalian 116029, People's Republic of China\\
$^{41}$ Liaoning University, Shenyang 110036, People's Republic of China\\
$^{42}$ Nanjing Normal University, Nanjing 210023, People's Republic of China\\
$^{43}$ Nanjing University, Nanjing 210093, People's Republic of China\\
$^{44}$ Nankai University, Tianjin 300071, People's Republic of China\\
$^{45}$ National Centre for Nuclear Research, Warsaw 02-093, Poland\\
$^{46}$ North China Electric Power University, Beijing 102206, People's Republic of China\\
$^{47}$ Peking University, Beijing 100871, People's Republic of China\\
$^{48}$ Qufu Normal University, Qufu 273165, People's Republic of China\\
$^{49}$ Renmin University of China, Beijing 100872, People's Republic of China\\
$^{50}$ Shandong Normal University, Jinan 250014, People's Republic of China\\
$^{51}$ Shandong University, Jinan 250100, People's Republic of China\\
$^{52}$ Shanghai Jiao Tong University, Shanghai 200240,  People's Republic of China\\
$^{53}$ Shanxi Normal University, Linfen 041004, People's Republic of China\\
$^{54}$ Shanxi University, Taiyuan 030006, People's Republic of China\\
$^{55}$ Sichuan University, Chengdu 610064, People's Republic of China\\
$^{56}$ Soochow University, Suzhou 215006, People's Republic of China\\
$^{57}$ South China Normal University, Guangzhou 510006, People's Republic of China\\
$^{58}$ Southeast University, Nanjing 211100, People's Republic of China\\
$^{59}$ State Key Laboratory of Particle Detection and Electronics, Beijing 100049, Hefei 230026, People's Republic of China\\
$^{60}$ Sun Yat-Sen University, Guangzhou 510275, People's Republic of China\\
$^{61}$ Suranaree University of Technology, University Avenue 111, Nakhon Ratchasima 30000, Thailand\\
$^{62}$ Tsinghua University, Beijing 100084, People's Republic of China\\
$^{63}$ Turkish Accelerator Center Particle Factory Group, (A)Istinye University, 34010, Istanbul, Turkey; (B)Near East University, Nicosia, North Cyprus, 99138, Mersin 10, Turkey\\
$^{64}$ University of Bristol, H H Wills Physics Laboratory, Tyndall Avenue, Bristol, BS8 1TL, UK\\
$^{65}$ University of Chinese Academy of Sciences, Beijing 100049, People's Republic of China\\
$^{66}$ University of Groningen, NL-9747 AA Groningen, The Netherlands\\
$^{67}$ University of Hawaii, Honolulu, Hawaii 96822, USA\\
$^{68}$ University of Jinan, Jinan 250022, People's Republic of China\\
$^{69}$ University of Manchester, Oxford Road, Manchester, M13 9PL, United Kingdom\\
$^{70}$ University of Muenster, Wilhelm-Klemm-Strasse 9, 48149 Muenster, Germany\\
$^{71}$ University of Oxford, Keble Road, Oxford OX13RH, United Kingdom\\
$^{72}$ University of Science and Technology Liaoning, Anshan 114051, People's Republic of China\\
$^{73}$ University of Science and Technology of China, Hefei 230026, People's Republic of China\\
$^{74}$ University of South China, Hengyang 421001, People's Republic of China\\
$^{75}$ University of the Punjab, Lahore-54590, Pakistan\\
$^{76}$ University of Turin and INFN, (A)University of Turin, I-10125, Turin, Italy; (B)University of Eastern Piedmont, I-15121, Alessandria, Italy; (C)INFN, I-10125, Turin, Italy\\
$^{77}$ Uppsala University, Box 516, SE-75120 Uppsala, Sweden\\
$^{78}$ Wuhan University, Wuhan 430072, People's Republic of China\\
$^{79}$ Yantai University, Yantai 264005, People's Republic of China\\
$^{80}$ Yunnan University, Kunming 650500, People's Republic of China\\
$^{81}$ Zhejiang University, Hangzhou 310027, People's Republic of China\\
$^{82}$ Zhengzhou University, Zhengzhou 450001, People's Republic of China\\
}
\vspace{0.2cm}{
$^{a}$ Deceased\\
$^{b}$ Also at the Moscow Institute of Physics and Technology, Moscow 141700, Russia\\
$^{c}$ Also at the Novosibirsk State University, Novosibirsk, 630090, Russia\\
$^{d}$ Also at the NRC "Kurchatov Institute", PNPI, 188300, Gatchina, Russia\\
$^{e}$ Also at Goethe University Frankfurt, 60323 Frankfurt am Main, Germany\\
$^{f}$ Also at Key Laboratory for Particle Physics, Astrophysics and Cosmology, Ministry of Education; Shanghai Key Laboratory for Particle Physics and Cosmology; Institute of Nuclear and Particle Physics, Shanghai 200240, People's Republic of China\\
$^{g}$ Also at Key Laboratory of Nuclear Physics and Ion-beam Application (MOE) and Institute of Modern Physics, Fudan University, Shanghai 200443, People's Republic of China\\
$^{h}$ Also at State Key Laboratory of Nuclear Physics and Technology, Peking University, Beijing 100871, People's Republic of China\\
$^{i}$ Also at School of Physics and Electronics, Hunan University, Changsha 410082, China\\
$^{j}$ Also at Guangdong Provincial Key Laboratory of Nuclear Science, Institute of Quantum Matter, South China Normal University, Guangzhou 510006, China\\
$^{k}$ Also at MOE Frontiers Science Center for Rare Isotopes, Lanzhou University, Lanzhou 730000, People's Republic of China\\
$^{l}$ Also at Lanzhou Center for Theoretical Physics, Lanzhou University, Lanzhou 730000, People's Republic of China\\
$^{m}$ Also at the Department of Mathematical Sciences, IBA, Karachi 75270, Pakistan\\
$^{n}$ Also at Ecole Polytechnique Federale de Lausanne (EPFL), CH-1015 Lausanne, Switzerland\\
$^{o}$ Also at Helmholtz Institute Mainz, Staudinger Weg 18, D-55099 Mainz, Germany\\
$^{p}$ Also at Hangzhou Institute for Advanced Study, University of Chinese Academy of Sciences, Hangzhou 310024, China\\
}
}

\begin{abstract}
Using $e^+e^-$ collision data collected with the BESIII detector
operating at the Beijing electron positron collider, the cross section
of $e^+e^-\to \pi^+\pi^- h_c$ is measured at 59 points with center-of-mass energy
$\sqrt{s}$ ranging from $4.009$ to $4.950~\mathrm{GeV}$ with a total
integrated luminosity of $22.2~\mathrm{fb}^{-1}$. The cross section
between $4.3$ and $4.45~\mathrm{GeV}$ exhibits a plateaulike shape
and drops sharply around $4.5~\mathrm{GeV}$, which cannot be described
by two resonances only.
Three coherent Breit-Wigner functions are used to parametrize the $\sqrt{s}$-dependent cross section line shape. The masses and widths
are determined to be
$M_1=(4223.6_{-3.7-2.9}^{+3.6+2.6})~\mathrm{MeV}/c^2$,
$\Gamma_1=(58.5_{-11.4-6.5}^{+10.8+6.7})~\mathrm{MeV}$,
$M_2=(4327.4_{-18.8-9.3}^{+20.1+10.7})~\mathrm{MeV}/c^2$,
$\Gamma_2=(244.1_{-27.1-18.3}^{+34.0+24.2})~\mathrm{MeV}$,
and $M_3=(4467.4_{-5.4-2.7}^{+7.2+3.2})~\mathrm{MeV}/c^2$,
and $\Gamma_3=(62.8_{-14.4-7.0}^{+19.2+9.9})~\mathrm{MeV}$.
The first uncertainties are statistical and the second are systematic.
The inclusion of the relatively narrower 
third component proves crucial for 
reproducing the drop at around 4.5~GeV.
The statistical significance of the three-resonance
assumption over the two-resonance assumption is greater than $5\sigma$.

\end{abstract}

\maketitle

The study of vector charmoniumlike states ($\jpc=1^{--}$, known as
$Y$ states) has generated significant interest.
The overpopulation of these $Y$ states
have led to exotic interpretations,
including
hybrid~\cite{Zhu:2005hp, Close:2005iz, Kou:2005gt,Brambilla:2022hhi,Oncala:2017hop},
tetraquark~\cite{Maiani:2014aja,Wang:2021qus}, molecule~\cite{Cleven:2013mka, Ding:2008gr, Wang:2013cya,Chen:2017abq} and
hadrocharmonium states~\cite{Dubynskiy:2008mq, Li:2013ssa},
or kinematically induced peaks~\cite{Chen:2017uof}. 
Meanwhile, the possibility that these states are excited charmonium states cannot be 
completely ruled out~\cite{Wang:2019mhs,Chen:2018fsi,Godfrey:1985xj}.
According to calculations based on an unquenched potential model,
the $4S-3D$ and $5S-4D$ mixing charmonium states are predicted to lie
between $4.2$ and $4.5~\text{GeV}/c^2$, with widths ranging from
$30$ to $80~\text{MeV}$~\cite{Wang:2019mhs},
the $\psi(4230)$, $\psi(4360)$, $\psi(4415)$, and $\psi(4500)$ are assigned
to be these states. 
Precise measurement of their properties is essential to unraveling their nature.

\begin{figure*}[htbp]
\begin{center}
\includegraphics[width=0.35\textwidth]{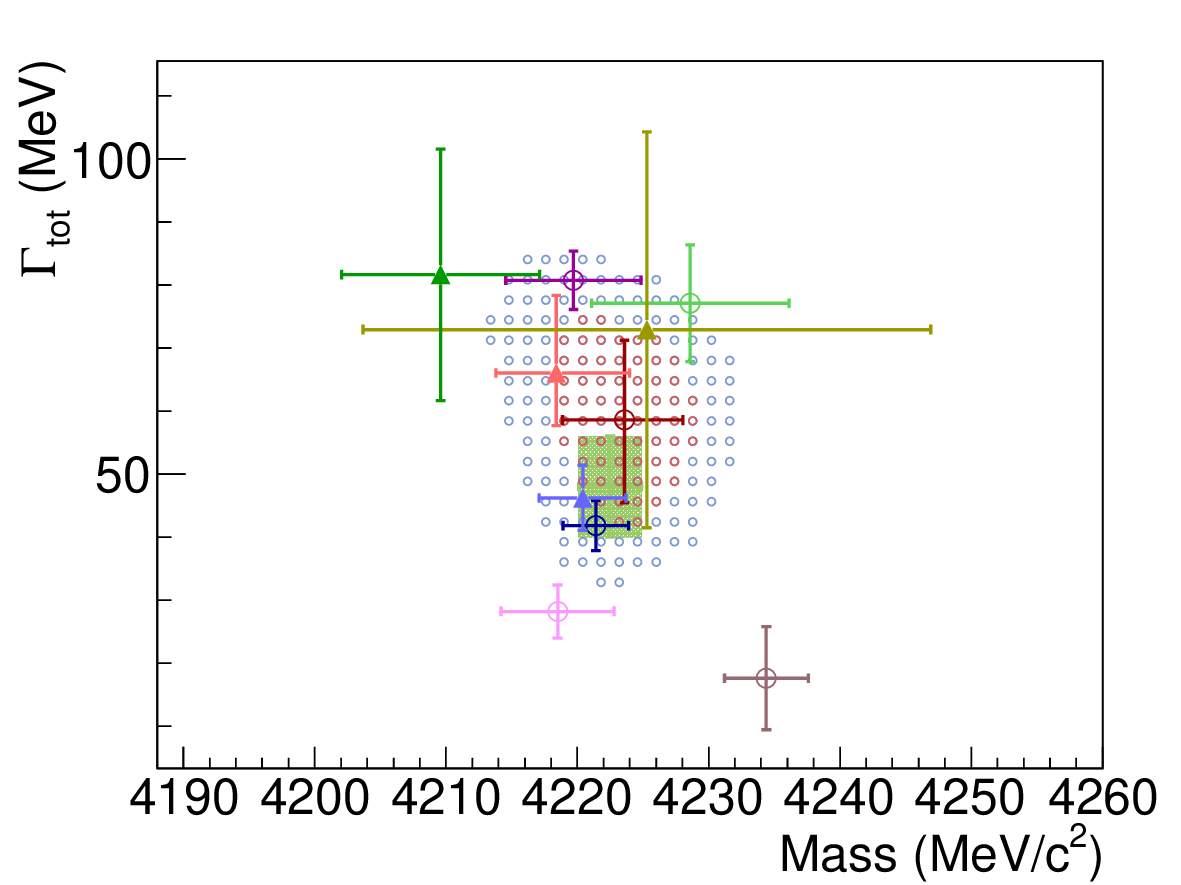}
\includegraphics[width=0.35\textwidth]{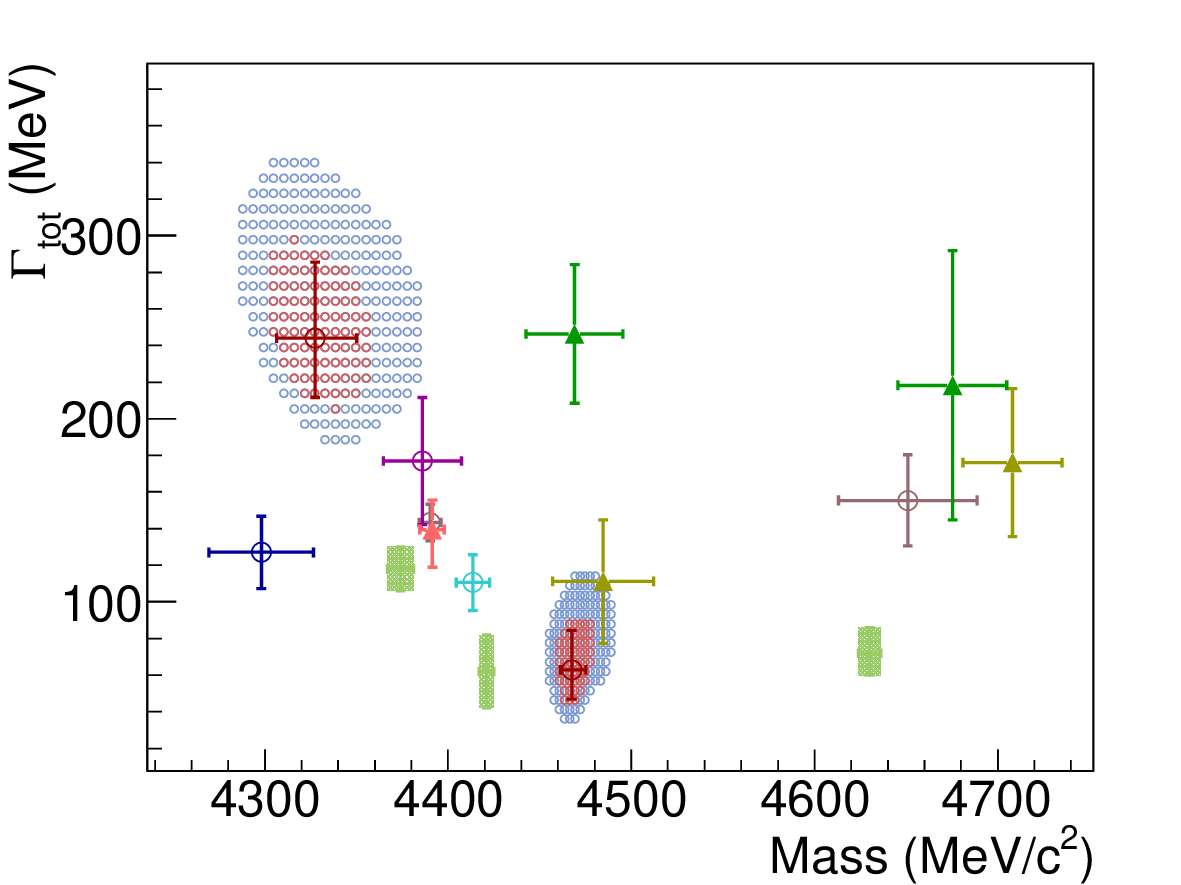}
\includegraphics[width=0.25\textwidth]{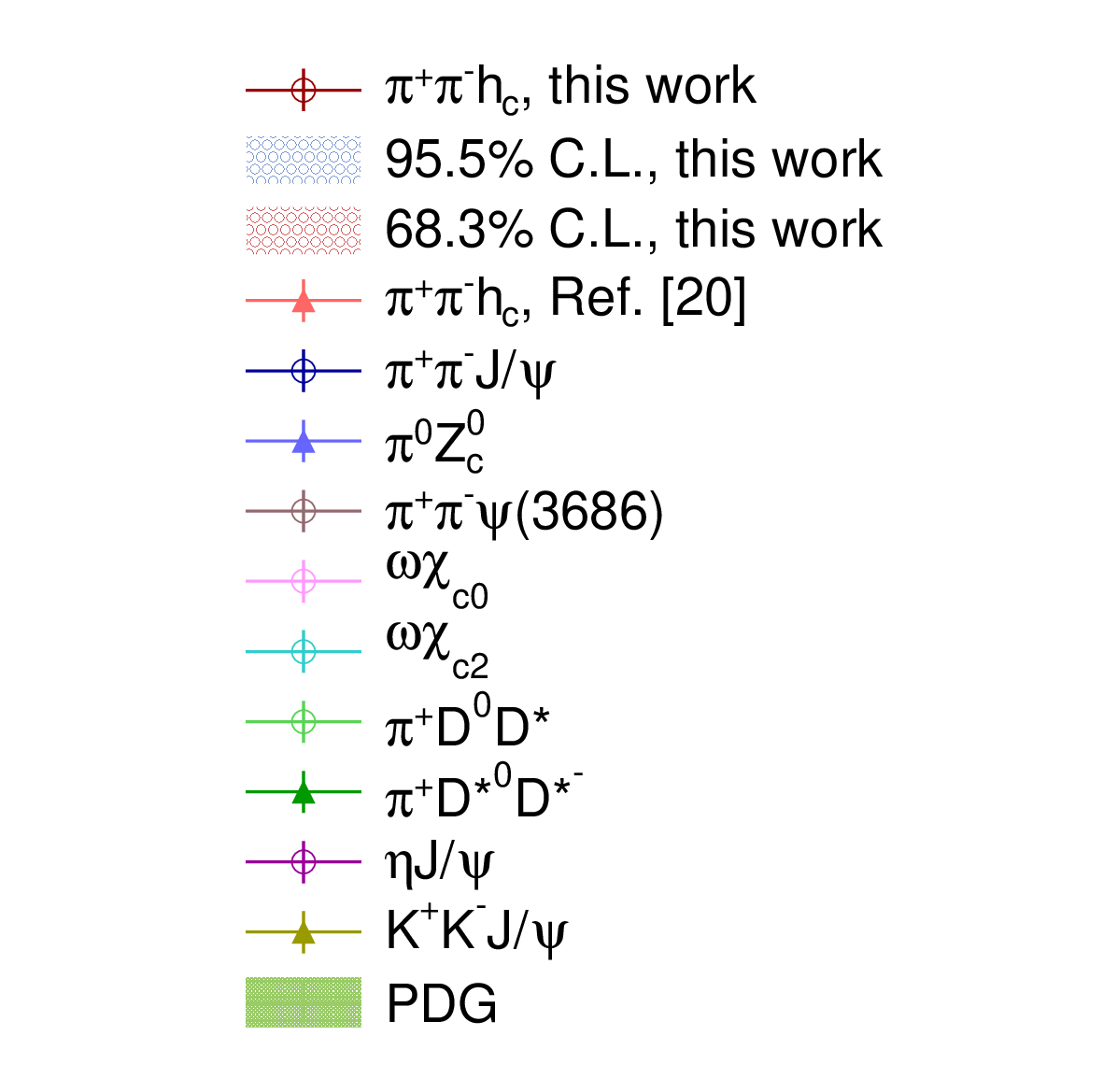}
\end{center}
\caption{Comparison of resonance parameters from hidden-charm or
open-charm processes~\cite{BESIII-pipihc,BESIII:pipijpsi,BESIII:pi0Zc0,BESIII:pipipsip,BESIII:omegaChic0,BESIII:omegaChic12,BESIII:piDDstar,BESIII:piDstarDstar,BESIII:etaJpsi,BESIII:KKJpsi1,BESIII:KKJpsi2}
(represented by dots with error bars in various colors),
along with the parameters of
$\psi(4230)$, $\psi(4360)$, $\psi(4415)$, and $\psi(4660)$ from PDG~\cite{PDG:2024}
(indicated by the green shaded square).
The red and blue shaded regions represent the 68.3\% and 95.5\% CL
regions determined in this work.
}
\label{fig:Res_comparison}
\end{figure*}

Among the processes in which the $Y$ states are observed, those containing
$\hc$ in the final state are particularly interesting. This is because
transitions between vector charmonium states and $\hc$ are expected to be suppressed
due to heavy quark spin symmetry, so a strong coupling is indicative of an exotic
internal structure, such as hybrid
configurations~\cite{Chen:2016ejo,BRAMBILLA20201}.
The $\EE\to\pp\hc$ process was first observed by the CLEO Collaboration
at a center-of-mass (c.m.) energy $\sqrt{s}=4.17~\gev$~\cite{CLEO-hcpaper}.
Subsequently, the BESIII experiment studied the $\EE\to\pp\hc$ cross section
with $\sqrt{s}$ ranging from $3.896$ to $4.600~\gev$ and observed the $Y(4220)$ and $Y(4390)$~\cite{BESIII-pipihc}.
Figure~\ref{fig:Res_comparison}
presents the resonance parameters of $Y(4220)$ and $Y(4390)$ alongside those
obtained from other
processes~\cite{BESIII:pipijpsi,BESIII:pi0Zc0,BESIII:pipipsip,BESIII:omegaChic0,BESIII:omegaChic12,BESIII:piDDstar,BESIII:piDstarDstar,BESIII:etaJpsi,BESIII:KKJpsi1,BESIII:KKJpsi2},
based on the BESIII scan samples.
In the $Y(4390)$ region,
resonances observed in different processes
show significant variation in mass and width.
At higher energies,
new vector structures around $4.75~\gev$ have been reported by BESIII in
$\EE\to K\bar{K}\jpsi$~\cite{BESIII:KKJpsi1,BESIII:KKJpsi2} and
$\EE\to D_{s}^{*}D_{s}^{*}$~\cite{BESIII:DssDss} processes. The decays
of these higher $Y$ states to $\hc$ have not been investigated yet.

In this Letter, we report a
measurement of the $\EE\to\pp h_c$
cross section at $\sqrt{s}$ from $4.009$ to $4.950~\gev$. The data
are collected with the BESIII detector~\cite{Ablikim:2009aa} and include
three sets: 19 energy points with large statistics ~\cite{BESIII-pipihc} (referred to as XYZ-I),
25 energy points with lower statistics (referred to as XYZ-II),
and 15 energy points, each with statistics of $8~{\rm pb}^{-1}$
(referred to as R-scan).
The integrated luminosity of these samples is $22.2~{\rm fb}^{-1}$,
determined from large-angle
Bhabha events with an uncertainty of $1\%$~\cite{BESIII-lumi-yifan, cms-lumi-round1314}.
The c.m.~energies for the XYZ-I and XYZ-II samples are determined from
$\EE\to\MM$ or $\EE\to\Lambda_c\bar{\Lambda}_c$ events~\cite{cms-offline-XYZ1,cms-offline-XYZ2, cms-lumi-round1314},
those for the R-scan samples are measured using multihadron final
states.

In this study, the $h_c$ is reconstructed via its electric-dipole
transition $\hc\to\gamma\etac$ with $\etac\to X_{i}$, where
$X_{i}$ signifies 16 exclusive hadronic final states: $\ppbar$,
$\pipipipi$, $\kkkk$, $\pipikk$, $\pipippbar$, $\pipipipipipi$,
$\pipipipikk$, $\kskpi$, $\kskpipipi$, $\kkpiz$, $\ppbarpiz$,
$\kketa$, $\pipieta$, $\pipipipieta$, $\pipipizpiz$, and
$\pipipipipizpiz$.
The $\ks$ is reconstructed using its decay to $\pp$, while $\piz$
and $\eta$ are reconstructed through their $\gamma\gamma$ final state.

Monte Carlo (MC) samples are used to determine the detection
efficiencies and to estimate the background contributions.
They are produced with a {\sc geant4}-based~\cite{geant4}
simulation software package, which includes the geometric description of the
BESIII detector and the detector response. The simulation
models the beam energy spread and initial state radiation (ISR)
in the $e^+e^-$ annihilations
with the generator {\sc kkmc}~\cite{ref:kkmc}. The maximum
energy of the ISR photon for the $\EE\to\pp\hc$ process corresponds
to its kinematical threshold. The inclusive MC sample includes
the production of open-charm processes, the ISR production of vector
charmonium(-like) states, and the continuum processes. All particle
decays are modeled with {\sc evtgen}~\cite{ref:evtgen} using
branching fractions either taken from Particle Data Group~\cite{PDG:2024},
when available, or otherwise estimated with
{\sc lundcharm}~\cite{ref:lundcharm}. Final state radiation
from charged final state particles is incorporated using the
{\sc photos} package~\cite{photos2}.

\begin{figure}[htbp]
\begin{center}
\includegraphics[width=0.45\textwidth]{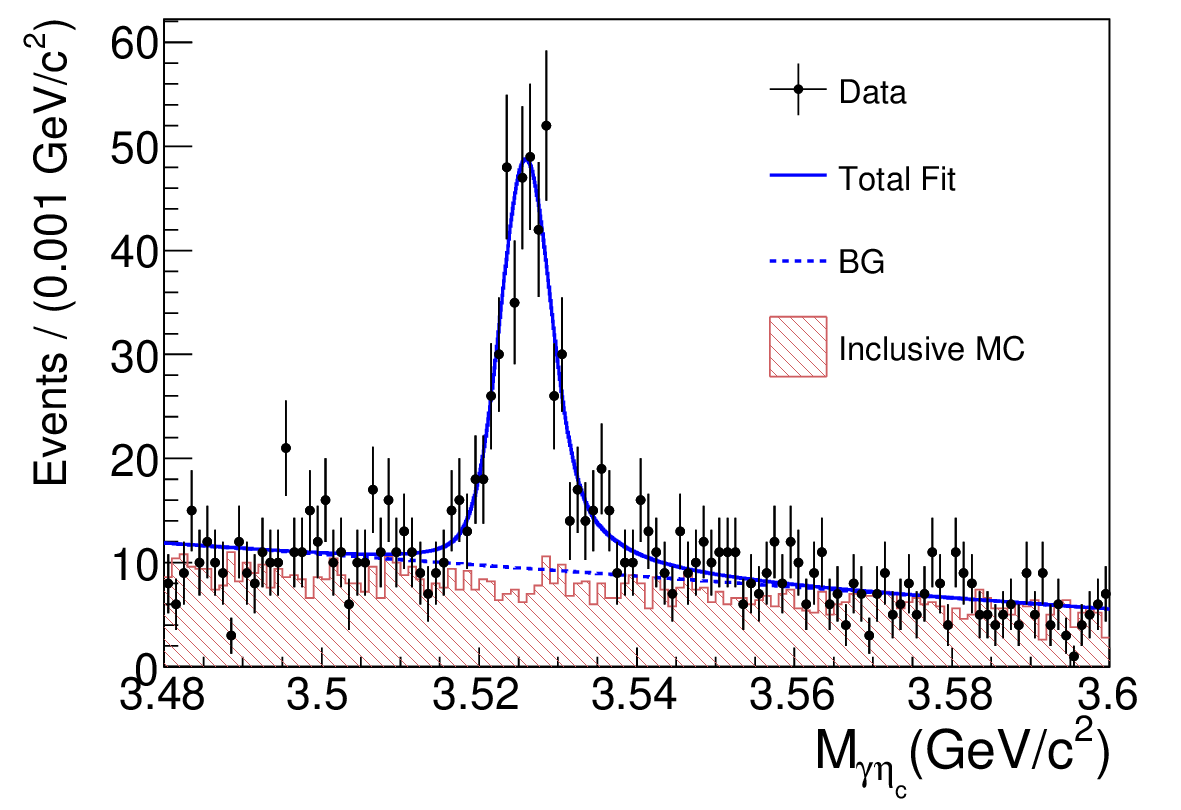}
\end{center}
\caption{The $M_{\gamma\etac}$ distribution in the $\etac$ signal
region at $\sqrt{s}=4.236~\gev$. Dots with error bars are the data,
the solid curve is the best fit result, and the dashed curve represents
the background contribution.
}
\label{fig:simu fit 4237}
\end{figure}

The event selection method is similar to the one used in Ref.~\cite{BESIII-pipihc}.
However, the mass windows of $\piz$, $\eta$, and $\etac$ and the requirement of
$\chi^2_{\rm 4C}$ are re-optimized to enhance the signal-to-background
ratio.
For the $\etac\to\pipipizpiz$ and $\etac\to\pipipipipizpiz$ modes,
we further require $\chi^2_{\rm {4C}}<\chi^2_{\rm {4C, \pm\gamma}}$,
where $\chi^2_{\rm {4C}}$ is taken from a four-constraint (4C) kinematic
fit of all selected final state particles with respect to the initial
$\EE$ four momentum, and $\chi^{2}_{\rm 4C, \pm\gamma}$ is taken from the
4C kinematic fit that includes or excludes one photon.
Figure~\ref{fig:simu fit 4237} shows the invariant mass distribution of
$\gamma\etac$ ($M_{\gamma\etac}$)
in the $\etac$ signal region
for the sum of the 16 decay channels at $\sqrt{s}=4.236~\gev$.
A clear $\hc\to\gamma\etac$ signal is observed. The background events
are distributed linearly in the $M_{\gamma\etac}$ distribution,
in agreement with the analysis of the inclusive MC sample.

The $\EE\to\pp\hc$ signal events yield is determined by performing
an unbinned maximum likelihood fit to the $M_{\gamma\etac}$ spectrum.
The signal contribution is modeled using the MC simulated shape,
convolved with a Gaussian function which accounts for the resolution difference
between data and the MC simulation. The background contribution is described
by a linear function. For the XYZ-I data sample, a
simultaneous fit to the 16 $\etac$ decay modes is performed. The numbers
of signal events in each mode are constrained according to
the detection efficiencies and branching fractions.
For the XYZ-II data sample, the $M_{\gamma\etac}$ spectra summed over
the 16 $\etac$ decay modes are fitted (referred to as the summed fit).
Additionally, the parameters of the Gaussian function are fixed to the
average values obtained from the fits to the XYZ-I data sample.
The consistency between the results from the simultaneous fit
and the summed fit is confirmed with the XYZ-I data sample.
For the R-scan data sample, the summed fit method is used,
with the background shape fixed according to that obtained from
the summed fit to the XYZ-I and XYZ-II data sample in the range
$4.3~\gev<\sqrt{s}<4.8~\gev$.

The Born cross section $\sigma^{\rm{Born}}$ is calculated via:
\begin{equation}\label{func:dressed cs}
\footnotesize
\frac{N^{\rm{obs}}}
{\mathcal{L}\cdot(1+\delta)\cdot(1/|1-\Pi|^2)
\cdot{\cal B}(\hc\to\gamma\etac)
\cdot\Sigma_{i=1}^{16}\epsilon_i{\cal B}(\etac\to X_i)},
\end{equation}
where $N^{\rm{obs}}$, $\cal{L}$, $(1+\delta)$, and $(1/|1-\Pi|^2)$ are
the signal yields, the integrated luminosity, the ISR correction
factor, and the vacuum polarization correction factor, respectively.
For the $i$th $\etac$ decay mode,
$\epsilon_{i}$ represents the detection efficiency,
and $\mathcal{B}(\etac\to X_i)$ denotes the branching fraction.
The branching fractions of $\hc\to\gamma\etac$ and the $\etac$ decays
are taken from previous BESIII measurements~\cite{Ablikim:2012ur,BESIII:2022tfo}.
The cross section is obtained through an iterative procedure,
since both the ISR factor and the detection efficiency 
depend on the cross section line shape~\cite{Sun:2020ehv}.
The procedure adopts the dressed cross section as its input,
which is the product of the Born cross section
and the vacuum polarization correction factor.
The dressed cross sections are shown in Fig.~\ref{fig:cs}
and summarized in the Supplemental
Material~\cite{SuppMat}, together with all the inputs used in the
calculation.

The $\sqrt{s}$-dependent dressed cross section is fitted using a maximum
likelihood method to investigate the vector resonance structures.
The best fit is achieved with a model incorporating three coherent
Breit-Wigner (BW) functions (the baseline model),
which is written as
\begin{equation}\label{func:3bw}
|BW_{1}(\sqrt{s}) + e^{i\phi_{2}}BW_{2}(\sqrt{s}) + e^{i\phi_{3}}BW_{3}(\sqrt{s})|^{2}.
\end{equation}
Here, $BW_{k}$ with $k=1,~2$ or $3$ is used to describe the resonance, defined as
\begin{equation}
\frac{M_{k}}{\sqrt{s}}\cdot \frac{\sqrt{12 \pi (\Gamma_{ee} \mathcal{B}(R_{k}\to\pp\hc))_{k} \Gamma_{k}}}{s-M_{k}^{2}+i M_{k}\Gamma_{k}}
\cdot\sqrt{\frac{PS(\sqrt{s})}{PS(M_{k})}}.
\end{equation}
In the fit, the mass $M_{k}$, the total width $\Gamma_{k}$, the product of
the electromagnetic width and the branching fraction
$[\Gamma_{ee}\mathcal{B}(R_k \to\pp\hc)]_{k}$,
and the relative phase $\phi_{k}$ are free parameters.

In the baseline model, 
four solutions with two sets of parameters 
are found in accordance with
expectations~\cite{Bai:2019jrb}.
The fit results are shown in Fig.~\ref{fig:cs},
and the resonance parameters are listed in Table~\ref{tab:fit_3bw}.
The mass versus width plots for the three resonance structures,
along with the 68.3\% and 95.5\% confidence level (CL) contours
are shown in Fig.~\ref{fig:Res_comparison},
together with the resonance parameters of vector charmonium(-like)
states observed in other processes.
The parameters of the first resonance are consistent with
those reported for $Y(4220)$ by BESIII.
However, the mass and width of the second resonance 
[$M=(4327.4_{-18.8}^{+20.1})~\mevcc$,
$\Gamma_{\rm tot}=(244.1_{-27.1}^{+34.0})~\mev$]
are found to be $60~\mevcc$ lower and $100~\mev$ wider 
compared to the reported values for $Y(4390)$ from previous study
of the same process [$M=(4391.6\pm6.3\pm1.0)~\mevcc$,
$\Gamma_{\rm tot}=(139.5\pm16.1\pm0.6)~\mev$]~\cite{BESIII-pipihc,SuppMat}.
This discrepancy arises from the inclusion of a third resonance in this work.
The model using two coherent BW functions cannot describe the dip in 
the cross section at $\sqrt{s}= 4.498~\gev$~\cite{SuppMat}.
The fit quality is calculated to be $\chi^2/ndf = 77.9/66$
in the two-resonance fit and $41.9/70$ when the third resonance is added.
Result of the two-resonance fit is shown in
Supplemental Material~\cite{SuppMat}.
The statistical significance of the
third resonance is $5.4\sigma$, estimated by utilizing the changes in likelihood
values ($\delta(-2\ln L)=38.8$) and the number of degrees of freedom ($\delta(ndf)=4$).

Several parametrization models are tested.
Adding one resonance with free parameters or a phase space
term [$PS(\sqrt{s})/s^{n}$] to the baseline model slightly improves the
fit quality. 
The statistical significance of this fourth resonance (or phase
space term) is $0.7\sigma$ ($0.1\sigma$), 
where $PS(\sqrt{s})$ is the
three-body phase space factor.
The model used in Ref.~\cite{Chen:2017uof} is also tested,
yielding a nonconvergent fit.
The cross section is then updated
using the baseline model as input cross section line shape and iterated
until convergence.

\begin{figure}[htbp]
\begin{center}
\includegraphics[width=0.45\textwidth]{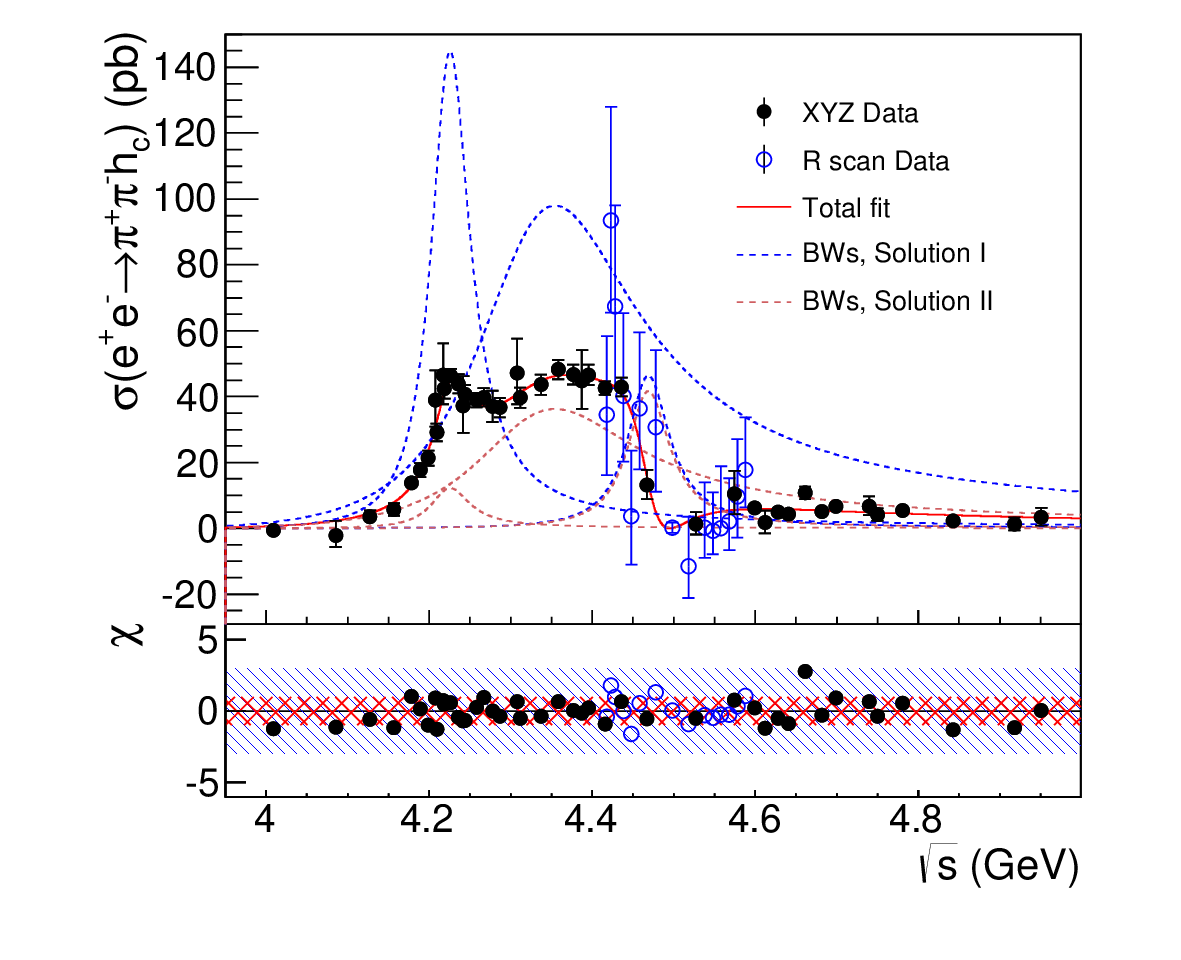}
\end{center}
\caption{
Fit to the dressed cross section for the two solutions for $\EE\to\pp\hc$ with
the baseline model (the red curve).
The blue and red dashed curves are contributions from the three structures.
The dots with error bars are the converged cross section.
The bottom panel shows the $\chi$ values,
in which the red and blue shadings
represent $\pm1$ and $\pm3$, respectively.
}
\label{fig:cs}
\end{figure}

\begin{table*}[htbp]
\begin{center}
\caption{
The fit results from the baseline model. The first uncertainty is statistical and second
systematic. The numbers in brackets
are from the second solution with equal fit quality.
}\label{tab:fit_3bw}

\footnotesize
\begin{tabular}{c|c|c|c}

\hline\hline
Parameter & $R_1$ & $R_2$ & $R_3$ \\
\hline
$M$ ($\mevcc$)
&$4223.6_{-3.7-2.9}^{+3.6+2.6}$
&$4327.4_{-18.8-9.3}^{+20.1+10.7}$
&$4467.4_{-5.4-2.7}^{+7.2+3.2}$
\\ 
$\Gamma$ (MeV)
&$58.5_{-11.4-6.5}^{+10.8+6.7}$
&$244.1_{-27.1-18.3}^{+34.0+24.2}$
&$62.8_{-14.4-7.0}^{+19.2+9.9}$
\\ 
$\Gamma_{ee}\mathcal\cdot{\cal B}(R\to\pp\hc)$ (eV)
&$10.2_{-1.5-1.4}^{+1.2+1.4}$ $(0.9_{-0.4-0.2}^{+0.4+0.3})$
&$29.1_{-3.9-3.4}^{+5.7+4.4}$ $(10.8_{-1.8-1.5}^{+2.5+1.9})$
&$3.9_{-1.7-0.5}^{+3.5+1.7}$ $(3.5_{-1.6-0.7}^{+3.0+1.5})$
\\ 
$\phi$~(rad)
& ...
& $3.6_{-0.1-0.1}^{+0.1+0.1}$ $(0.7 _{-0.3-0.2}^{+0.3+0.2})$
& $0.7_{-0.3-0.2}^{+0.3+0.1}$ $(-2.2_{-0.3-0.1}^{+0.3+0.2})$
\\ \hline
\end{tabular}
\end{center}
\end{table*}

Systematic uncertainties in the cross section measurement come mainly
from the integrated luminosity,
the statistical uncertainties of the c.m.~energy for the R-scan data sample,
the input cross section line shape,
the branching fractions,
the detection efficiency,
and the determination of $N^{\rm obs}$.
The uncertainty of the integrated luminosity is
$1\%$~\cite{BESIII-lumi-yifan, cms-lumi-round1314}.
The effect from the statistical uncertainties
of the c.m.~energy for the R-scan data sample
is estimated by shifting $\sqrt{s}$ by $\pm$1~MeV.
The uncertainty from the parametrization of the cross section line shape is
estimated by adding a phase space term to Eq.~\ref{func:3bw}.
The difference
is $1\%$, and is taken as the uncertainty.
The uncertainty of the cross section is
reflected by the uncertainty of the parameters in the formula
used to describe the cross section line shape.
This is estimated by sampling these parameters
according to
the covariance matrix and recalculating the ISR factor, and
the standard deviation of the resultant distribution is taken as
the systematic uncertainty.
The combined branching fractions ${\cal B}(\hc\to\gamma\etac)\cdot{\cal B}(\etac\to X_i)$
are taken from Ref.~\cite{Ablikim:2012ur}, updated with the latest
measurement of $\psi(3686)\to\pi^0\hc$ from BESIII~\cite{BESIII:2022tfo},
giving an uncertainty of $9.7\%$.

The uncertainty related to the detection efficiency contains
the tracking efficiency, photon reconstruction, $\ks$ reconstruction,
$\piz/\eta$ mass window, $\etac$ mass window, $\chi^2_{\rm 4C}$ requirement,
and intermediate states in $\pi\hc$ and $\pp$ system.
The first four terms are not added in this study since they are
included in the branching fraction of $\etac$.
The uncertainties of the two additional pion tracks accompanying the $\hc$
are also included ($1\%$ per track).
The uncertainties from the mass, width~\cite{PDG:2024}, and line
shape of $\etac$~\cite{Anashin:2010dh} used in MC simulations are
estimated by varying them within uncertainties or adding the missing
terms and check the difference in detection efficiencies, which
are $1.1\%$ for the $\etac$ parameters and $0.2\%$ for the line shape.
The uncertainty from the applied requirement on the $\chi^2_{\rm 4C}$
value is estimated by correcting the helix parameter of charged particles
to match the resolution in data~\cite{bes3-kinematicfit-eff}. The
uncertainty from the requirement of
$\chi^2_{\rm {4C}}<\chi^2_{\rm {4C, \pm\gamma}}$
is estimated by removing this requirement and repeating the analysis. The
systematic uncertainties for the two aforementioned terms are $2.1\%$
and $2.3\%$, respectively.
The uncertainty from the intermediate states in $\pi\hc$ and $\pp$
system is estimated by reweighting the MC simulation using a
Dalitz plot obtained from data, and is $8.0\%$, $12.5\%$, and
$3.5\%$ for data the samples at $\sqrt{s}=4.189~\gev$, $\sqrt{s}=4.199~\gev$,
and the other c.m.~energies.

The uncertainties in the determination of $N^{\rm obs}$ are estimated
by varying the fit conditions and observing the resulting changes
in the cross section results.
Uncertainties from the fixed parameters in the fit,
including the mass resolution difference between
data and MC simulation for XYZ-II and R-scan data sample,
as well as the background shape for R-scan data sample,
are estimated by adjusting each parameter by 1 standard deviation.
To access the uncertainty from the background shape,
the linear function is replaced with a second order Chebyshev function,
the impact on the results is negligible.
The uncertainty from the fit range is tested by modifying the nominal
fit range by $\pm5$ and $\pm10~\mevcc$ and examining the uncorrelated
uncertainty as outlined in~\cite{Barlow:2002yb, Barlow2},
which is found negligible.
The total systematic uncertainty in the
$\EE\to\pp\hc$ cross section measurement,
listed in Supplemental Material~\cite{SuppMat},
is determined by assuming these sources as independent.

The systematic uncertainties for the parameters of the resonance
structures are summarized in Table~\ref{tab:sys:Ys}.
They primarily arise from the systematic and statistical uncertainties
of the c.m.~energy, the beam energy spread, the systematic uncertainty of the
cross section, and the choice of the parametrization model.
The impact of the systematic uncertainty in the c.m.~energy measurement
is $0.6~\mev$~\cite{cms-offline-XYZ1,cms-offline-XYZ2,cms-lumi-round1314}
and only affects the mass measurements.
The effect from the statistical uncertainty in c.m.~energy measurement
for the R-scan data sample is estimated by randomly modifying
the corresponding $\sqrt{s}$ values according to a Gaussian function with mean $0$
and standard deviation $1~\mev$, and reevaluating the resonance parameters.

The uncertainties from cross section measurement are divided into
two classes. ``Cross section I" relates to the uncorrelated terms,
including the mass resolution difference between data and MC
simulation, fixed background shape for the R-scan sample,
ISR factor, and uncorrelated systematic uncertainty terms in
the detection efficiency
(the requirement of $\chi^2_{\rm {4C}}<\chi^2_{\rm {4C, \pm\gamma}}$
and the intermediate states in $\pi\pi\hc$ system).
They are considered by adding these terms to the statistical uncertainty.
The systematic uncertainty for each parameter is calculated
with $\sqrt{\delta_{\rm w/}^2-\delta_{\rm w/o}^2}$,
where $\delta_{\rm w/}$ and $\delta_{\rm w/o}$ are
the uncertainties with and without the systematic terms included.
``Cross section II" represents the correlated terms
common to all data samples, estimated to be $10.2\%$.
The uncertainty from the parameterization model is estimated by
adding a phase space term to the baseline model.
The cross section is obtained through an iterative procedure,
since both the ISR factor and the detection efficiency depend on
the cross section line shape~\cite{Sun:2020ehv}.
The systematic uncertainty due to iteration stability is the maximum 
difference (1.0\%) across all energy points between the final two iterations.
The uncertainty from the beam energy spread is estimated by convolving
a Gaussian function (with the standard deviation provided by the beam
energy measurement system~\cite{Abakumova:2011rp}) to the fit formula.

\begin{table*}[htbp]
\caption{The systematic uncertainty in the measurement
of resonance parameters of the $Y$ states.
The numbers in brackets indicate uncertainty of the
second solution.
Syst. and stat. refer to systematic and statistical uncertainties of the c.m. energy measurement,
respectively.
}\label{tab:sys:Ys}
\begin{center}
\footnotesize
\begin{tabular}{c|ccc|cccc|cccc}
\hline\hline
\multirow{3}{*}{Sources} &  \multicolumn{3}{c|}{$R_1$}
& \multicolumn{4}{c|}{$R_2$}  & \multicolumn{4}{c}{$R_3$}
\\
&  $M$   & $\Gamma_{\rm tot}$ & $\Gamma_{ee}\cdot\mathcal{B}$
&  $M$   & $\Gamma_{\rm tot}$ & $\Gamma_{ee}\cdot\mathcal{B}$ & $\phi$
&  $M$   & $\Gamma_{\rm tot}$ & $\Gamma_{ee}\cdot\mathcal{B}$ & $\phi$
\\
&  $(\mevcc)$   & $(\mev)$ & (\%)
&  $(\mevcc)$   & $(\mev)$ & (\%) & (rad)
&  $(\mevcc)$   & $(\mev)$ & (\%) & (rad)
\\
\hline
c.m. energy (syst.)        & 0.6 & ... & ...
& 0.6 & ... & ... & ...         & 0.6 & ... & ... & ... \\

c.m. energy (stat.)        & 0.0 & 0.1 & 0.0 (0.3)
& 0.1 & 0.4 & 0.2 (0.2) &0.0 (0.0)        & 0.1 & 0.3 & 0.8 (0.6) &0.0 (0.0) \\

Cross section I
& $^{+2.5}_{-2.8}$ & $^{+6.4}_{-6.1}$ & $^{+8.1}_{-8.0}$ ($^{28.2}_{17.9}$)
& $^{+10.6}_{-9.1}$ & $^{+20.2}_{-13.1}$ & $^{+10.3}_{-5.7}$ ($^{11.1}_{4.3}$)
& $^{+0.1}_{-0.1}$ ($^{0.2}_{0.2}$)
& $^{+2.9}_{-2.4}$ & $^{+8.9}_{-6.1}$ & $^{+37.8}_{-4.5}$ ($^{38.4}_{14.6}$)
& $^{+0.1}_{-0.2}$ ($^{0.2}_{0.1}$)\\

Cross section II & ... & ... & 10.2
& ... & ... & 10.2 & ...        & ... & ... & 10.2 & ...\\

Parametrization & 0.4 & 1.8 & 3.2 (9.4)
& 1.1 & 12.3 & 0.7 (8.6) & 0.0 (0.0)
& 0.8 & 2.4 & 3.9 (4.9) & 0.1 (0.0)\\

Energy spread
& $^{+0.4}_{-0.3}$ & $^{+1.1}_{-1.3}$ & $^{+0.8}_{-1.8}$ ($^{3.3}_{3.4}$)
& $^{+0.7}_{-1.1}$ & $^{+5.1}_{-3.5}$ & $^{+4.0}_{-1.1}$ ($^{4.7}_{1.8}$)
& $^{+0.0}_{-0.0}$ ($^{0.0}_{0.0}$)

& $^{+0.9}_{-0.7}$ & $^{+3.6}_{-2.3}$ & $^{+19.3}_{-1.7}$ ($^{17.7}_{1.2}$)
& $^{+0.0}_{-0.0}$ ($^{0.0}_{0.0}$)\\

\hline
Total &
$^{+2.6}_{-2.9}$ & $^{+6.7}_{-6.5}$ &
$^{+13.5}_{-13.5}$ ($^{31.6}_{22.9}$)
&
$^{+10.7}_{-9.3}$ & $^{+24.2}_{-18.3}$ &
$^{+15.1}_{-11.8}$ ($^{18.0}_{14.2}$) &
$^{+0.1}_{-0.1}$ ($^{0.2}_{0.2}$)
&
$^{+3.2}_{-2.7}$ &  $^{+9.9}_{-7.0}$ &
$^{+43.8}_{-12.0}$ ($^{43.8}_{18.6}$) &
$^{+0.1}_{-0.2}$ ($^{0.2}_{0.1}$)
\\
\hline
\end{tabular}

\end{center}
\end{table*}

In summary, we measure the $\EE\to\pp\hc$~cross section at 59 energy points from 
$\sqrt{s}=4.009$~to $4.951~\gev$. The cross section between $4.3$~and~$4.45~\gev$
exhibits a plateaulike shape and has a dip at $4.5~\gev$. 
The best description of the cross section line shape is achieved
by the coherent sum of three BW functions. 
The significance of the third resonance is larger than $5\sigma$.
No obvious resonance structure is observed at around $\psi(4660)$, which is in tension 
with the theoretical prediction in a hidden charm $P$-wave tetraquark model~\cite{Ali:2017wsf}.

The mass and width of the first resonance are consistent with the $\psi(4230)$~\cite{PDG:2024}
and the observation in a previous study of the same process~\cite{BESIII-pipihc}.
The mass of the second resonance is consistent with the $\psi(4360)$~\cite{PDG:2024},
but the obtained width is about $100~\mev$ broader.
It is noteworthy that the mass of the second resonance is much closer
to the resonance observed in $\EE\to\pp\jpsi$~\cite{BESIII:pipijpsi}, with respect
to the previous study~\cite{BESIII-pipihc}.
The parameters of the third resonance are consistent with the $\psi(4500)$
found in $\kk\jpsi$~\cite{BESIII:KKJpsi1,BESIII:KKJpsi2},
whereas the mass is $40~\mev$ higher than the $\psi(4415)$.

The model proposed in Ref.~\cite{Chen:2017uof} cannot describe the cross
section line shape, where the structure around $4.39~\gev$
is attributed to the interference between $\psi(4160)$ and $\psi(4415)$.
Subsequent studies predict two pairs of $S-D$ mixing vector charmonium 
states~\cite{Wang:2019mhs}.
The masses of $R_1$ and $R_2$ align with the $4S-3D$ mixing model;
however, the width of $R_2$ significantly exceeds the predicted limit of
$\Gamma_2 \leq 80\mev$.
While $R_3$ could be one of the $5S-4D$ states, its expected partner is not seen.
Additionally, the mass of $R_2$ and $R_3$ are also close to that of $\psi(3D)$,
yet the large width of $R_2$ is incompatible with the model~\cite{Chen:2018fsi,Godfrey:1985xj}.
Notably, the mass and width of $R_{3}$ are consistent with a hybrid state prediction~\cite{Brambilla:2022hhi}.

Reference~\cite{Cao:2020vab} suggests
$\mathcal{O}(10^2)~{\rm
eV}\lesssim\Gamma^{Y(4260)}_{ee}\lesssim\mathcal{O}(10^3)~{\rm eV}$.
Assuming 
$\Gamma^{R_1,R_2}_{ee} \in (10^2, 10^3)~{\rm eV}$,
we determine $\Gamma^{R_1}_{\pip\pim\hc} \in (0.05,0.5)$~MeV or $(0.6,6.0)$~MeV and
$\Gamma^{R_2}_{\pip\pim\hc} \in (2.6,26.4)$~MeV or $(7.1,71.0)$~MeV for the two solutions.
$\Gamma^{R_1}_{\pp\hc}$ lies within the upper limit 
$\Gamma^{\psi(4230)}_{\pp h_c}<1.26~\mev$
set by a molecular model calculation\cite{Chen:2017abq}.
The $\Gamma^{R_1}_{\pip\pim\hc}$ is smaller than hybrid configuration predictions,
which are $\Gamma^{\psi(4230)}_{h_c+l.h.}=17(15)$
and $\Gamma^{\psi(4360)}_{h_c+l.h.}=14(12)~\mev$,
where $l.h.$ stands for light hadrons~\cite{Oncala:2017hop}.
However, the model cannot be excluded due to large uncertainties of the theoretical
result.

$Acknowledgments-$The BESIII Collaboration thanks the staff of BEPCII~\cite{BEPC link} and the IHEP computing center for their strong support. This work is supported in part by National Key R\&D Program of China under Contracts No. 2020YFA0406300, No. 2020YFA0406400, No. 2023YFA1606000, No. 2023YFA1606704; National Natural Science Foundation of China (NSFC) under Contracts No. 12375070, No. 11635010, No. 11935015, No. 11935016, No. 11935018, No. 12025502, No. 12035009, No. 12035013, No. 12061131003, No. 12192260, No. 12192261, No. 12192262, No. 12192263, No. 12192264, No. 12192265, No. 12221005, No. 12225509, No. 12235017, No. 12361141819; the Chinese Academy of Sciences (CAS) Large-Scale Scientific Facility Program; CAS under Contract No. YSBR-101;
Joint Large-Scale Scientific Facility Funds of the NSFC and CAS under Contract No.U2032108; Shanghai Leading Talent Program of Eastern Talent Plan under Contract No. JLH5913002;
100 Talents Program of CAS; The Institute of Nuclear and Particle Physics (INPAC) and Shanghai Key Laboratory for Particle Physics and Cosmology; Agencia Nacional de Investigación y Desarrollo de Chile (ANID), Chile under Contract No. ANID PIA/APOYO AFB230003; German Research Foundation DFG under Contract No. FOR5327; Istituto Nazionale di Fisica Nucleare, Italy; Knut and Alice Wallenberg Foundation under Contracts No. 2021.0174, No. 2021.0299; Ministry of Development of Turkey under Contract No. DPT2006K-120470; National Research Foundation of Korea under Contract No. NRF-2022R1A2C1092335; National Science and Technology fund of Mongolia; National Science Research and Innovation Fund (NSRF) via the Program Management Unit for Human Resources \& Institutional Development, Research and Innovation of Thailand under Contract No. B50G670107; Polish National Science Centre under Contract No. 2024/53/B/ST2/00975; Swedish Research Council under Contract No. 2019.04595; U. S. Department of Energy under Contract No. DE-FG02-05ER41374

$Data~availability-$The data 
that support the findings of this article are openly available~\cite{HepData}.

\bibliography{basename of .bib file}

\end{document}